\documentclass[12pt,preprint]{aastex}

\shorttitle{Dwarf Nova Outbursts in Nova Her 1960}
\shortauthors{Honeycutt, Robertson \& Kafka}

\begin{document}


\title{The Dwarf Nova Outbursts of Nova Her 1960 (=V446~Her)}


\author{R.K. Honeycutt\altaffilmark{1}, J.W. Robertson\altaffilmark{2}, 
S. Kafka\altaffilmark{3}}

\altaffiltext{1}{Astronomy Department, Indiana University, Swain Hall West,
 Bloomington, IN 47405. E-mail: honey@astro.indiana.edu}

\altaffiltext{2}{Arkansas Tech University, Dept. of Physical Sciences,
 1701 N. Boulder Ave., Russellville, AR 72801-2222. E-mail: jrobertson@atu.edu}

\altaffiltext{3}{Dept.of Terrestrial Magnetism, Carnegie Inst. of Washington,
 5241 Broad Branch Road NW, Washington, DC 2001. E-mail: skafka@dtm.ciw.edu}

\begin{abstract}

V446~Her is the best example of an old nova which has developed dwarf nova
eruptions in the post-nova state.  We report on observed properties of the
long-term light curve of V446~Her, using photometry
over 19 years.  Yearly averages of the outburst magnitudes shows a decline 
of $\sim$0.013 mag yr$^{-1}$, consistent with the decline of other 
post-novae that do not have dwarf nova outbursts.  Previous suggestions of 
bimodal distributions of the amplitudes and widths of the outbursts are 
confirmed.  The outbursts occur at a mean spacing of 18 days but the range 
of spacings is large (13-30 days).  From simulations of dwarf nova outbursts
it has been predicted that the outburst spacing in V446~Her will 
increase as $\dot{M}$ from the red dwarf companion slowly falls following 
the nova; however the large intrinsic scatter in the spacings serves to
hide any evidence of this effect.  We do find a systematic change in
the outburst pattern in which the brighter, wider type of outbursts 
disappeared after late 2003, and this phenomenon is suggested to be due
to falling $\dot{M}$ following the nova.

\end{abstract}

\keywords{stars:individual(V446Her)--novae,cataclysmic variables}

\section{Introduction}

V446~Her was a rather undistinguished nova, reaching V$\sim$6 in 1960.
It is at $\sim$ 1 kpc distance and has an orbital period of 4.97 h 
(Thorstensen \& Taylor 2000).
Honeycutt et~al. (1995) (Paper I) reported that V446~Her had regular 
$\sim$1.5 mag outbursts (OBs) in monitoring data beginning 3 decades after
the nova eruption.  The OB amplitudes were smaller than those of dwarf novae 
(DN), leading to the tentative conclusion that the events were due to 
variations in mass transfer from the red dwarf (RD).

In a later paper Honeycutt et~al. (1998) (Paper II) found that V446~Her 
has two close neighboring stars (likely to be optical companions) whose 
constant light is usually blended 
with the light of V446~Her, attenuating the apparent amplitude of the
OBs.  When this effect is taken into account the OB amplitude
becomes $\sim$2.5 mag, in line with that expected from a DN.  
Paper II suggested that bi-modal distributions of OB spacings and 
OB amplitudes were present
in V446~Her, concluding that the OBs were indeed due to V446~Her 
having become a DN some decades after the novae.  Many old novae are declining 
in brightness by $\sim$0.01 mag yr$^{-1}$ 50 years after the nova event 
(Duerbeck 1992).  This
is consistent with the hibernation scenario of cyclic nova evolution
(Shara 1989) in which this decline in $\dot{M}$ eventually takes the 
accretion disk (AD) through the DN regime, and finally into hibernation.
Prior to the discovery of regular OBs in V446~Her the observational 
evidence for old nova becoming DN some decades after
the nova had been only modest (Livio 1987; 1990), but the V446~Her DN OBs
are a secure example of the phenomenon.  As such, these OBs
are a potentially sensitive test of AD theory and simulations,
in which the effect of slowly lowering $\dot{M}$ might be observed.

An additional 11 observing seasons of V446~Her photometry have accumulated 
since Paper II.  This new study was motivated by the possibility that a 
systematic evolution of the DN OB spacing might be found, as predicted by
AD simulations of the dependence of OB spacing on $\dot{M}$ (e.g. Ichikawa
\& Osaki 1994).  We 
will also examine other systematics of the V446~Her light curve OBs over 
nearly two decades, seeking confirmation of the bimodal OB properties which
were suggested in Paper II. 
  
\section{Data Acquisition and Reduction}

Our long-term photometry of V446~Her consists of two sets 1991-2004
and 2007-2009.  The first is from an unattended, autonomous 0.41-m telescope 
in central Indiana, informally called RoboScope (Honeycutt et~al. 1994 and 
references therein).  Flats and other detector calibration data were 
automatically acquired and applied each night, followed by aperture photometry 
and field identification, all using custom software (Honeycutt \& Turner 1992).  
Final photometric reductions were done using the incomplete ensemble photometry 
routines contained in Astrovar, which is a custom 
package based on the technique described in Honeycutt (1992), but with the 
addition of a graphical user interface.  The RoboScope Astrovar solution used
111 ensemble stars, and the zeropoint was established to within 0.025 mag using 5 
secondary standards from Henden \& Honeycutt (1997).  The second long-term sequence was
acquired 2007-2009 using an unattended, autonomous 1.25-m telescope at the same site as
RoboScope.  These data were reduced using a custom batch pipeline consisting of 
IRAF\footnote{IRAF is distributed by the National Optical Astronomy
  Observatories, which are operated by the Association of Universities for
  Research in Astronomy, Inc., under cooperative agreement with the National
  Science Foundation.} 
routines for detector calibrations, followed by the application of 
SExtractor\footnote{SEextractor is a source detection and photomery package
  described by Bertin and Arnouts 1996.  It is available from 
  http://terapix.iap.fr/soft/sextractor/.}.  The light curves were then 
generated using Astrovar, employing a total of 228 ensemble comparison stars.
The zero point was determined to within 0.022 mag using 9 secondary standards 
from Henden \& Honeycutt (1995).  The exposure times were 4-min for RoboScope, 
and 3-min for the 1.25-m; all exposures were in the V-band.  As in Papers I 
and II all magnitudes discussed here are for the blended light of V446~Her 
and the two close non-variable companions.

The number of useable RoboScope images is 1257, plus 127 from the 1.25-m 
telescope.  Figure 1 shows our full light curve.  The individual errors 
(which for clarity are not shown in the light curves) are for differential 
magnitudes with respect to the ensemble.  For the RoboScope data the average 
error in quiescence (taken to be V fainter than 16.5) is 0.06 mag, while the 
average error is 0.03 mag for V brighter than 16.5.  For the 1.25-m data the 
corresponding errors are 0.04 mag and 0.02 mag.  

The magnitude limit of RoboScope is V$\sim$18 on a clear moonless night, but 
this limit is degraded when moonlight and/or haze is present.  We see in 
Figure 1 that V446~Her does not fall to the magnitude limit of RoboScope on 
any night, some of which are clear and moonless, always remaining brighter 
than V$\sim$17.5.  However, when the magnitude limit is degraded, V446~Her 
is often not detected while in quiescence.  This means that the RoboScope 
data in quiescence are more poorly sampled than the brighter portions of the 
light curve, by nearly 2$\times$.  This effect is much less pronounced for 
the 1.25-m data.

Figures 2-7 show the light curve of V446~Her by observing season.  Some of 
the seasons are too poorly sampled to allow characterization of the 
individual OBs, but several seasons easily resolve sequences of 
OBs.

Table 1 lists the JDs, magnitudes, errors and telescope used.  The complete 
version of Table 1 is available only in electronic form; the data will also be 
archived with the AAVSO.

\section{Analysis and Discussion}

Decades after the nova event, many old novae are becoming fainter 
on average by $\sim$0.01 mag yr$^{-1}$ (Duerbeck 1992).  This is probably 
due to fading irradiation of the RD by the white dwarf (WD) 
as the WD  cools from the nova explosion, thereby decreasing
$\dot{M}$ and the accretion luminosity.  The same effects are expected 
to be present if the cataclysmic variable (CV) has become a DN.  However, unlike 
in a nova-like (NL) CV, 
the disk in a DN is not in steady-state so it is unclear what kind of 
photometric measure might be appropriate to test for this fading in
V446~Her.  In Figure 8 we plot the seasonal averages of the V446~Her 
data for 3 magnitude ranges.  The quiescence data show no systematic 
fading, which might be expected because the disk is far from steady-state
and is not expected to be reflecting $\dot{M}$ from the RD.
However, as seen in the top panel of Figure 8 the OB magnitudes show a fading of 
0.010$\pm$0.003 mag yr$^{-1}$.  When corrected for the contributions of the two
constant companion stars whose light is included in the measurements, the decline
becomes 0.013$\pm$0.003 mag yr$^{-1}$, still consistent with the Duerbeck value 
for old novae.    Because OB is the state in which most of the accretion luminosity 
is released; we take this as confirmation of the expected fading.  When 
we use all the data (middle curve in Figure 8) we again see no fading, 
but these data are dominated by the larger number of points in
quiescence.

Next we investigated the spacing, widths, and peak brightnesses of the 
individual OBs in V446~Her.  Many of the OBs are missed due to unfavorable 
data sampling, and even for detected OBs the spacing can be measured 
only if there are enough quiescence data between adjacent OBs to insure 
that no intervening OBs were missed.  Similarly the amplitudes and widths 
can be measured only if the light curve is reasonably well-sampled near the top
and sides respectively of the OB.  The OB parameters that are plotted and discussed 
are therefore for mostly (but not fully) the same set of OBs.

Ichikawa \& Osaki (1994) found from 1-D numerical simulations that the OB
recurrence interval scales as $\dot{M}^{-2}$ for 
$\dot{M}\gtrsim$10$^{16.3}$g s$^{-1}$,
below which the recurrence interval is constant.  Their simulations used 
system parameters appropriate to U Gem, and two treatments of $\alpha_{cold}$ 
were employed.  In Case A $\alpha_{cold}$ rises weakly with disk radius,
while in Case B it is held constant.  For Case A the recurrence interval 
becomes constant at 200 d while for Case B the recurrence interval becomes 
constant at 18 days.  For reference, the average OB recurrence interval in 
V446~Her is 18 d.  If we assume for the moment that V446~Her is always in the 
t$_{recur}\sim\dot{M}^{-2}$ regime, then a V-band fading of 1\% yr$^{-1}$ would 
result in a 2\% increase in t$_{recur}$ each year.  This would be easily 
discernable over 1-2 decades if it were not for the fact that in V446~Her 
t$_{recur}$ varies erratically over the range 13-30 days.
 
In Figure 9 we plot the spacing of adjacent OBs for a 14 year interval.
Although there are clearly systematic effects in this plot, there is no obvious 
monotonic behavior.  The reason for this may lie in the complicated patterns of 
OB behaviors, which may reflect a situation in which the effects of 
falling $\dot{M}$ on the OBs show up more in the character of the OBs 
rather than in the spacing.

In Paper II it was suggested that the OBs in V446~Her were bimodal in both 
amplitude and width.  This new larger data set can be used to confirm and 
expand upon these results.  In Figure 10 the distribution of amplitudes is 
clearly bimodal, as can also be seen in the light curves of Figures 2-7.  
Referring to the lower panel of Figure 10 it appears that the OB widths are 
also bimodal.
It is therefore not surprising that Figure 11 shows OB widths correlated with
the amplitudes.  The data points in Figure 11 seem to fall into two groupings 
corresponding to the double peaks in Figure 10.

In Figure 1 we see that data points brighter than 15.7 are missing after 
JD=2453000.  These missing data belong to the brighter, wider outubursts 
that make up longward tails of the distributions in Figure 10.  Might this
effect be due to small-number statistics?  This can be tested by
comparing the fraction of bright points before and after 2453000.  We wish to
exclude quiescence data in this analysis because the completeness
of points fainter than 16.5 changes when the data in Figure 1 switches from
the 0.41-m telescope to the 1.25-m near 2453500 (see $\S$2).  Restricting ourselves to
OB data (V$<$16.5) the ratio of the number of OB points brighter and fainter
than 15.7 is 76/503 = 0.15 for JD$<$2453000.  There are 102 data points
after 2453000.  If the same distribution of OB magnitudes holds we
expect 0.15(102) = 15 points brighter than 15.7 after 2453000, whereas we 
find zero. For a Poisson distribution with a mean of 15, the probability 
of finding zero is only 3$\times$10$^{-7}$.  However, this result is not
rigorous and is too optimistic (by an unknown but likely large amount)
because choosing a magnitude threshold of 15.7 before testing turns this 
into a {\em post hoc} analysis.  The Poisson test nevertheless suggests
that the effect is real, and we note that the missing brighter
OBs occur in data from both telescopes.  Only time will tell if this 
apparent change in outburst pattern is temporary or is 
part of a monotonic change.  We suspect that the missing brighter, wider OBs 
are the reaction of the accretion disk (AD) to falling $\dot{M}$,
and is the manner in which the AD releases less energy, as opposed to larger 
OB spacings.

Complex patterns of alternating OB amplitudes and shapes are common in some
DN such as SS Cyg (Cannizzo \& Mattei 1992; Cannizzo 1993) and may be 
triggered by variations in $\dot{M}$ from the RD 
(Schreiber, G\"{a}nsicke \& Hessman 2000a).  Such patterns are also common 
(and are in fact a defining characteristic) in SU UMa-type DN 
(O'Donoghue 2000; Hellier 2001).  SU UMa superoutbursts (SOBs) are
brighter and wider than the normal OBs in SU UMa systems, and are thought
to occur when the disk grows to a size sufficient for the outer portions of the
AD to orbit in a 3:1 resonance with the RD, leading to increased tidal torques
and occasional SOBs (Osaki 1996).  

Some authors have considered whether enhanced mass transfer from the RD during OB 
might account for the varying widths and amplitudes of DN OBs (e.g. Smak 1991;
Ichikawa, Hirose \& Osaki 1993; Smak 1999).  Hameury, Lasota \& Warner (2000)
investigated the combined effects of irradiation on both the accretion disk
and the secondary star, finding that the large variety of DN OB durations and
spacings may result from the interplay of these effects.  Such hydrid models
seem to be needed to explain some features of DN OB behavior, even if the tidal
instability mechanism remains valid for most SOB features.  Wide OBs which resulted
from the tidal instability would be expected to continue with falling $\dot{M}$
following the nova; it would simply take longer for the AD to reach the size needed 
for resonance.  On the other hand, wide OBs which resulted from enhanced $\dot{M}$ 
during OB might cease as the irradiation levels needed for this mechanism slowly 
falls following the nova.

Schreiber, G\"{a}nsicke \& Cannizzo (2000b) considered how the effects
of irradiation of the AD by the WD might inhibit DN OBs in post nova.  As the
WD in V446~Her continues to cool after the nova, they predict a decrease 
in OB frequency
and an increase in the amplitude of the OBs.  Any systematic change in OB 
spacing in V446~Her appears to be obscured by shorter-term changes in OB
spacing, perhaps related to chaotic changes in the shapes and strengths of 
the OBs, which might in turn be due to changes in
$\dot{M}$ from the RD.  We do not see the predicted increase in the amplitude
of the OBs in V446~Her. If anything, the average amplitude decreases with 
time because the larger amplitude, wider OBs become less common over 19 years.,
as seen by the absence of large amplitude outbursts during the last quarter
of the data in Figure 1.

\section{Conclusions}

The distributions of OB amplitudes and widths in V446~Her are found to 
be bimodal.  This is suggestive of either SS Cyg-type behavior (thought
to arise from the action of the thermal/viscous disk instability, perhaps 
aided by irradiation effects)
or from SU UMa-type OB/SOBs (thought to be due to combination of
the disk thermal limit cycle, plus a tidal instability in the
disk).  If a predicted systematic increase in the spacing of the OBs
is present, it is hidden by large intrinsic scatter in the recurrence intervals.

A systematic decline in mean brightness, as found in other post-nova
systems that do not have DN OBs, appears in V446~Her at 0.013 mag y$^{-1}$,
consistent with the observed decline in other old novae.  This
decline shows up only in the OB measures, probably because the
accretion luminosity reflecting declining $\dot{M}$ appears mostly
during OB.  

The brighter, wider-type OBs of V446~Her disappear abruptly after
JD 2453000.  If this turns out to be a long-term systematic
change in the OB pattern, then it appears to be consistent with  
enhanced $\dot{M}$ from the RD during DN OB as the mechanism responsible
for the bimodal distribution of the outburst widths.

It is somewhat disappointing that the firm prediction of increasing
recurrence interval for the V446~Her OBs cannot yet be measured in
V446 Her, even after $\sim$2 decades of photometric monitoring.
However, it is quite interesting that the very phenomena helping to
obscure this effect shows that, less than
3 decades after the nova, the V446 Her DN OBs already have 
many of the kinds of poorly-understood patterns of varying OB 
widths, spacings, and amplitudes seen in some other well-observed DN.
These outburst patterns therefore appear to be rather immune to any
lingering effects of the nova.

\begin{deluxetable}{cccc}
\tablenum{1}
\tablewidth{0pt}
\tablecolumns{4}
\tablecaption{Magnitudes for V446~Her}
\tablehead{\colhead{JD}  & \colhead{V Mag} & \colhead{Error} & \colhead{Source}}
\startdata
 2448425.71760  &  16.019  &  0.025 &  IU 0.41-m  \\
 2448427.79330  &  16.095  &  0.040 &  IU 0.41-m  \\
 2448444.72012  &  16.990  &  0.043 &  IU 0.41-m  \\
 2448453.83903  &  16.980  &  0.051 &  IU 0.41-m  \\
 2448454.65772  &  16.377  &  0.035 &  IU 0.41-m  \\
 2448454.75431  &  16.448  &  0.032 &  IU 0.41-m  \\
 2448504.60180  &  17.029  &  0.046 &  IU 0.41-m  \\
 2448506.56617  &  17.206  &  0.056 &  IU 0,41-m  \\
 2448511.55818  &  16.931  &  0.046 &  IU 0.41-m  \\
 2448528.52908  &  16.168  &  0.022 &  IU 0.41-m  \\
 2448536.54353  &  16.930  &  0.053 &  IU 0,41-m  \\
 2448542.60561  &  17.070  &  0.066 &  IU 0.41-m  \\
.............   & ......   & .....  &  .........  \\  
.............   & ......   & .....  & ..........  \\
.............   & ......   & .....  & ..........  \\
 2454972.75238  &  16.623  &  0.024 &  IU 1.25-m  \\
\enddata
\end{deluxetable}
\clearpage


\begin{figure}  
\epsscale{0.9}
\plotone{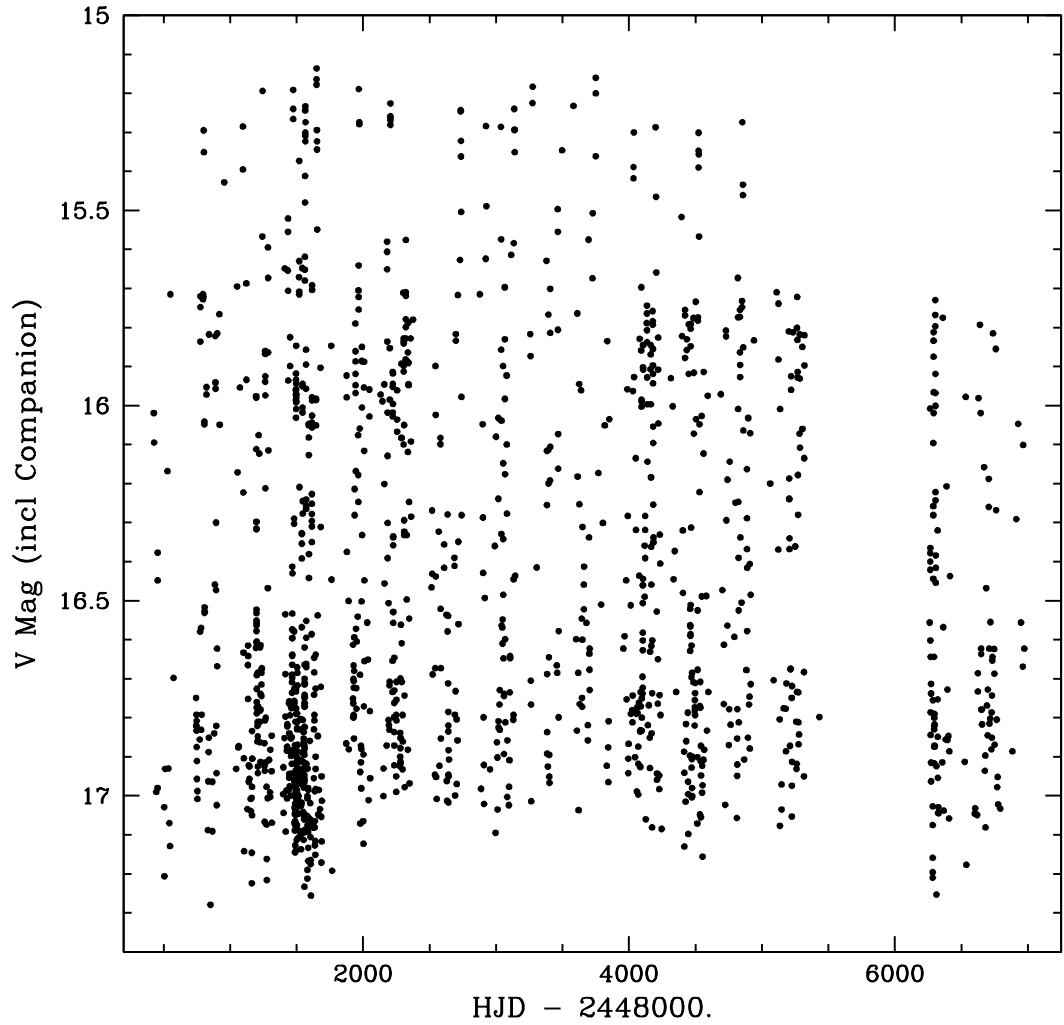}
\caption{The full light curve of V446~Her from 1991-Jun-18 to 2009-May-21.  The magnitudes
are for the blended light of V446~Her and its two close non-variable companion stars.} 
\end{figure}
\clearpage  

\begin{figure}  
\epsscale{0.9}
\plotone{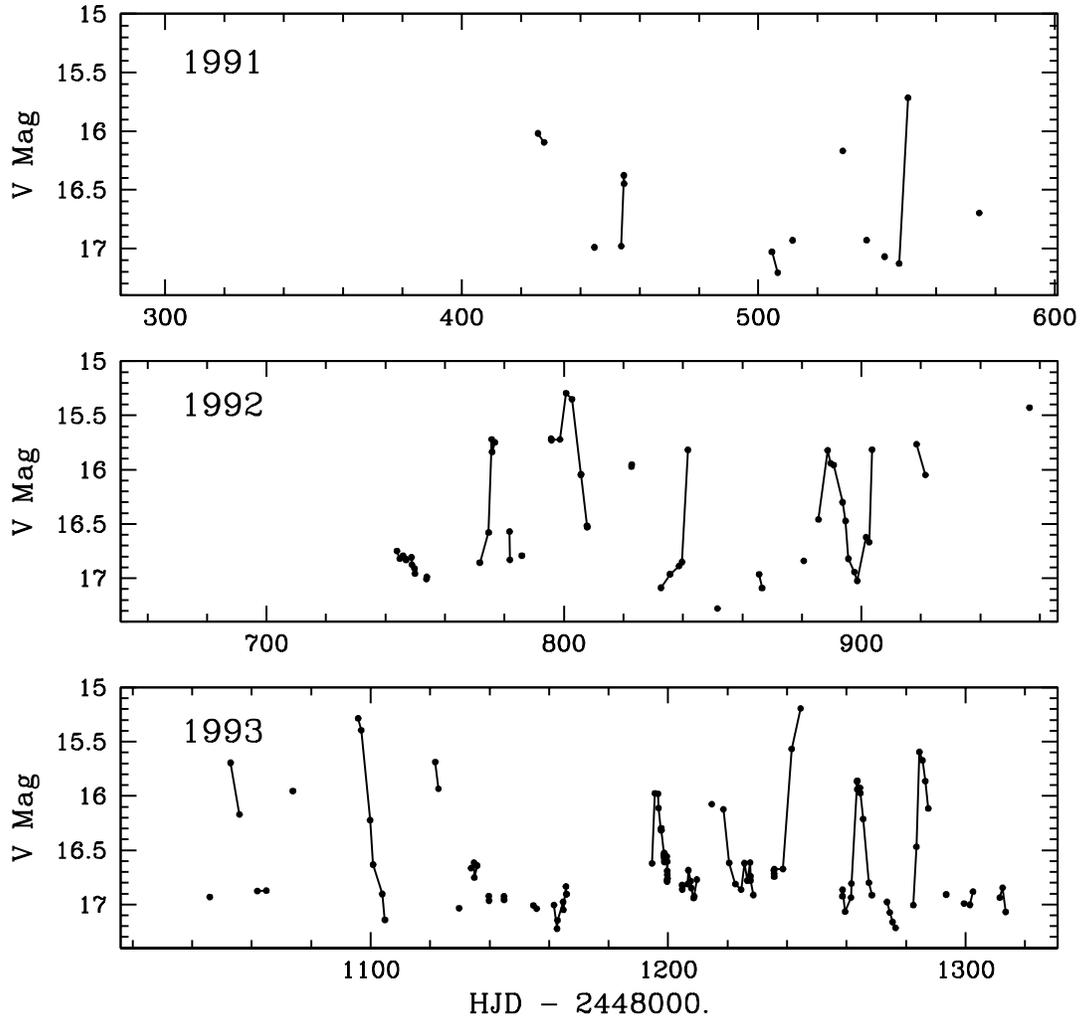}
\caption{Expanded portions of Figure 1 for 3 observing seasons 1991-1993,
using RoboScope data.  Data points separated by less than 3.5 days are 
connected with straight lines.} 
\end{figure}
\clearpage  

\begin{figure}  
\epsscale{0.9}
\plotone{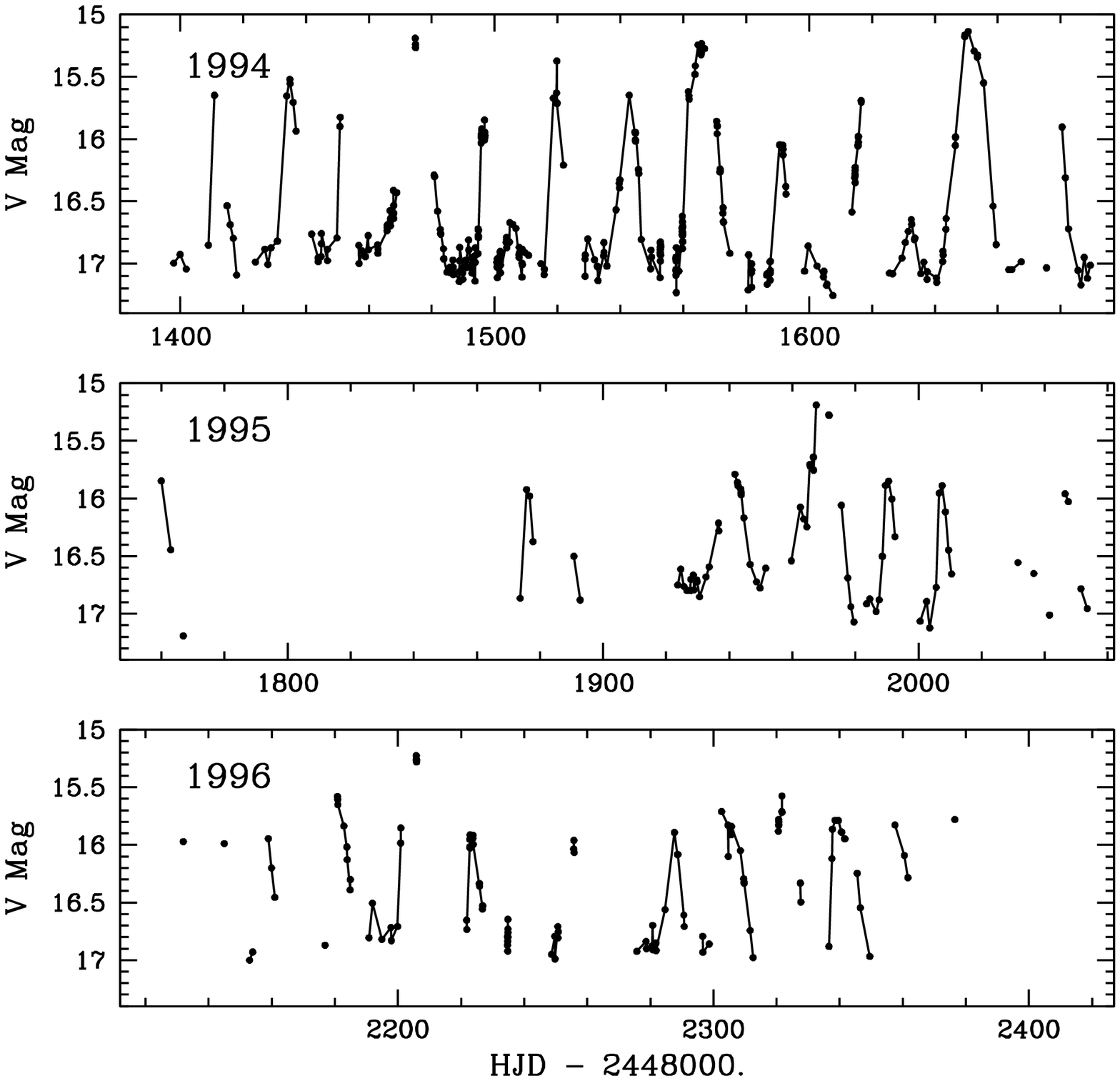}
\caption{Like Figure 2, except for observing seasons 1994-1996.} 
\end{figure}
\clearpage  

\begin{figure}  
\epsscale{0.9}
\plotone{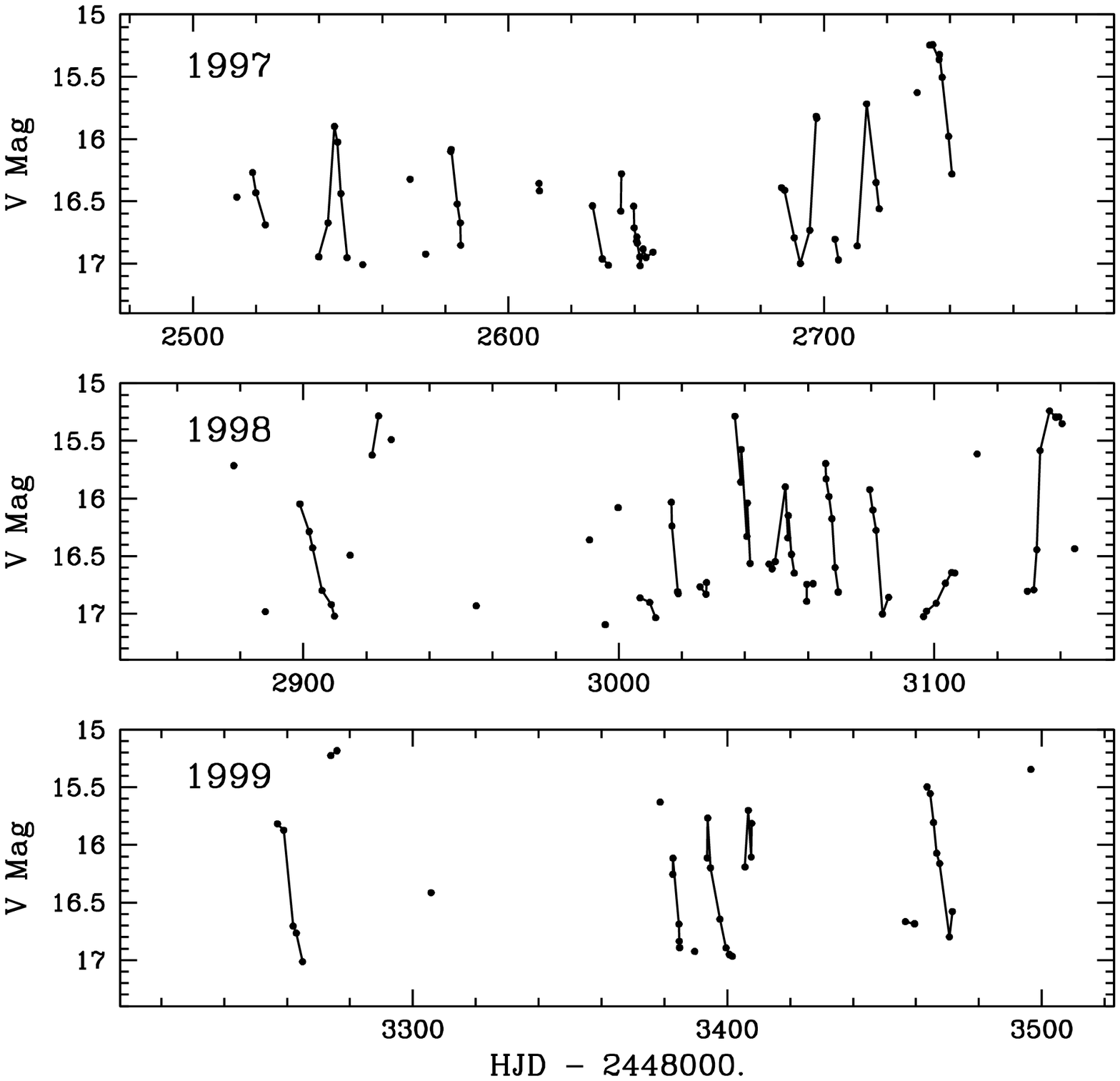}
\caption{Like Figure 2, except for observing seasons 1997-1999.} 
\end{figure}
\clearpage  

\begin{figure}  
\epsscale{0.9}
\plotone{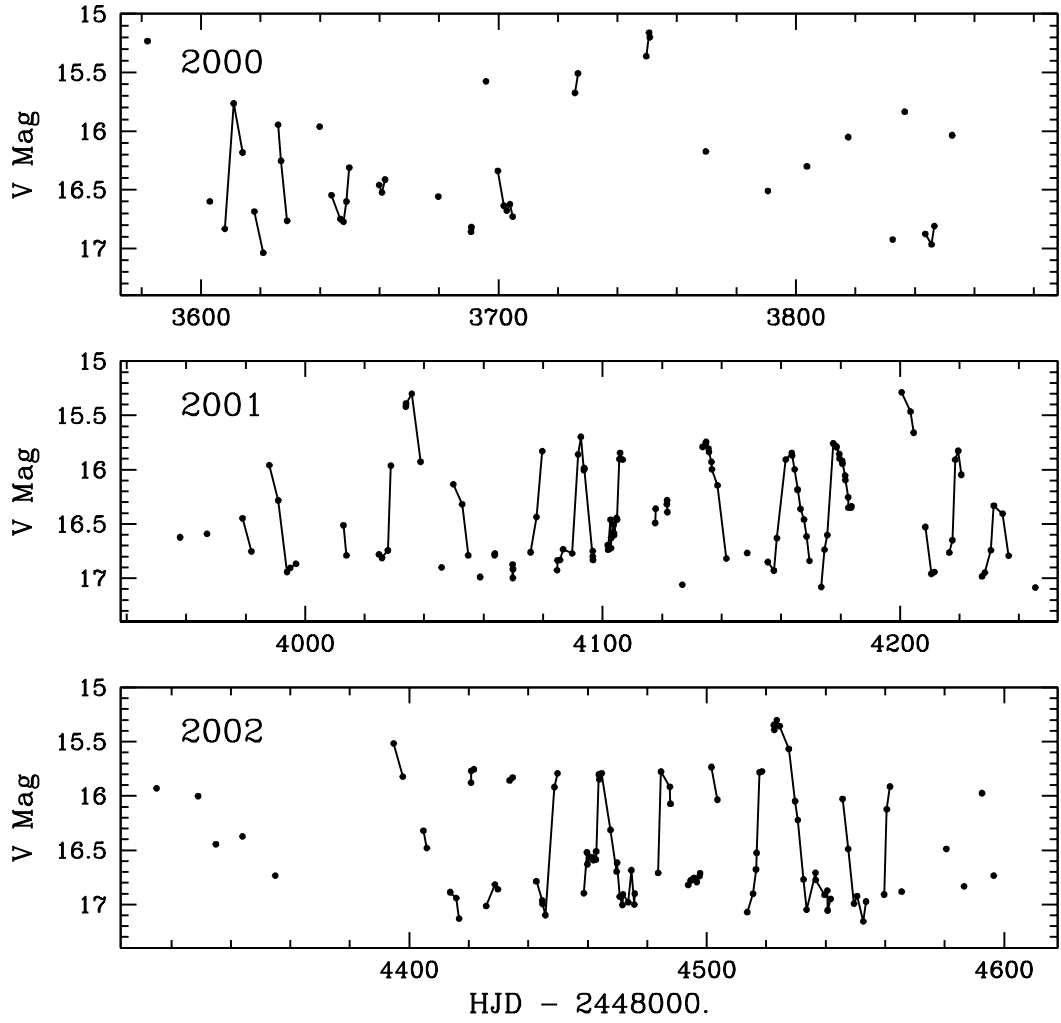}
\caption{Like Figure 2, except for observing seasons 2000-2002.} 
\end{figure}
\clearpage  

\begin{figure}  
\epsscale{0.9}
\plotone{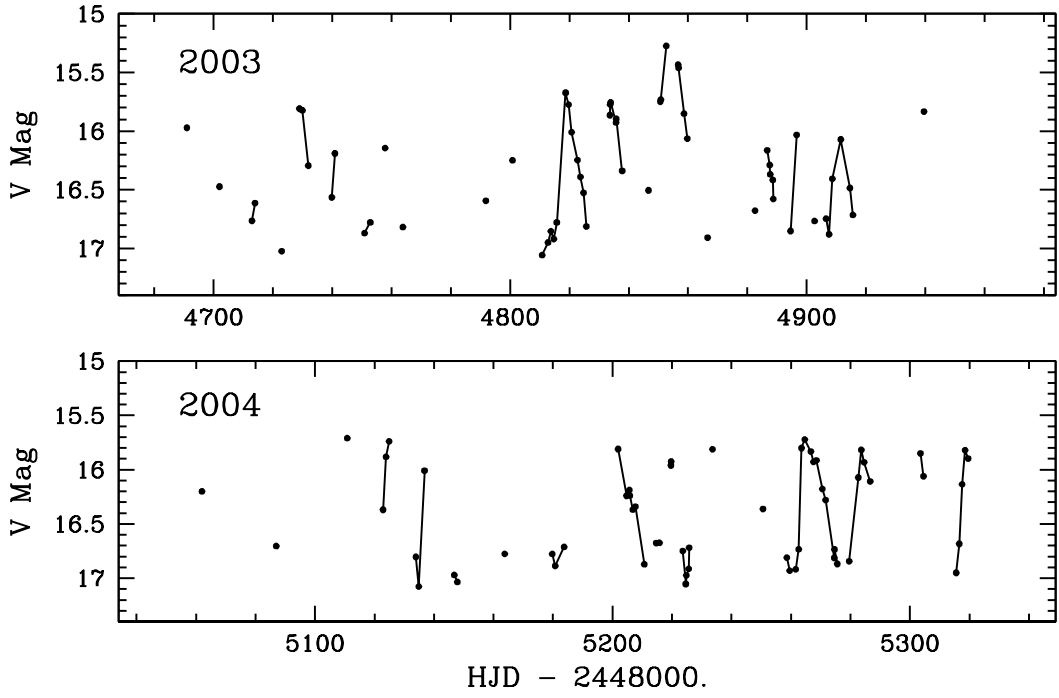}
\caption{Like Figure 2, except for observing seasons 2003 and 2004.} 
\end{figure}
\clearpage  

\begin{figure}  
\epsscale{0.9}
\plotone{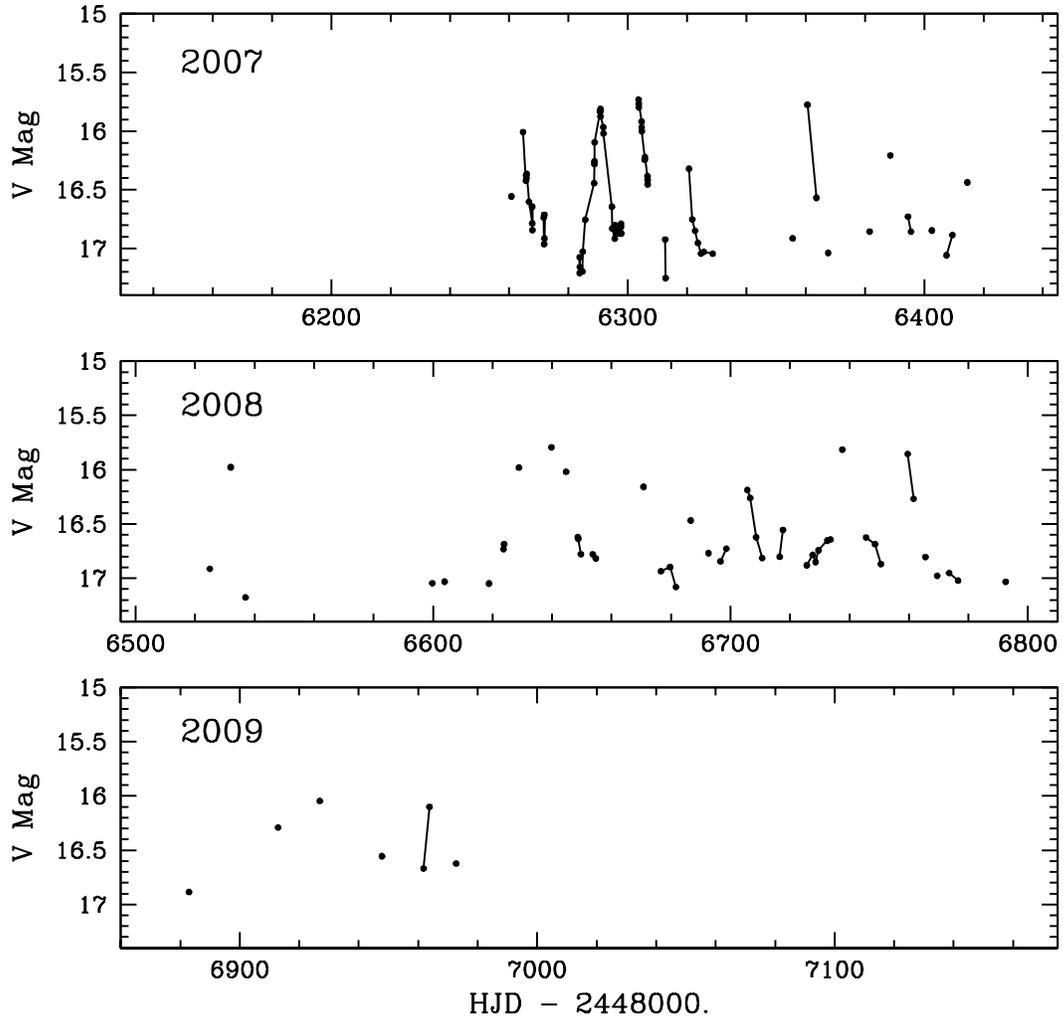}
\caption{Like Figure 2, except that these data for years 2007-2009 were
obtained using the Indiana Univ. 1.25-m telescope.} 
\end{figure}
\clearpage  

\begin{figure}  
\epsscale{0.9}
\plotone{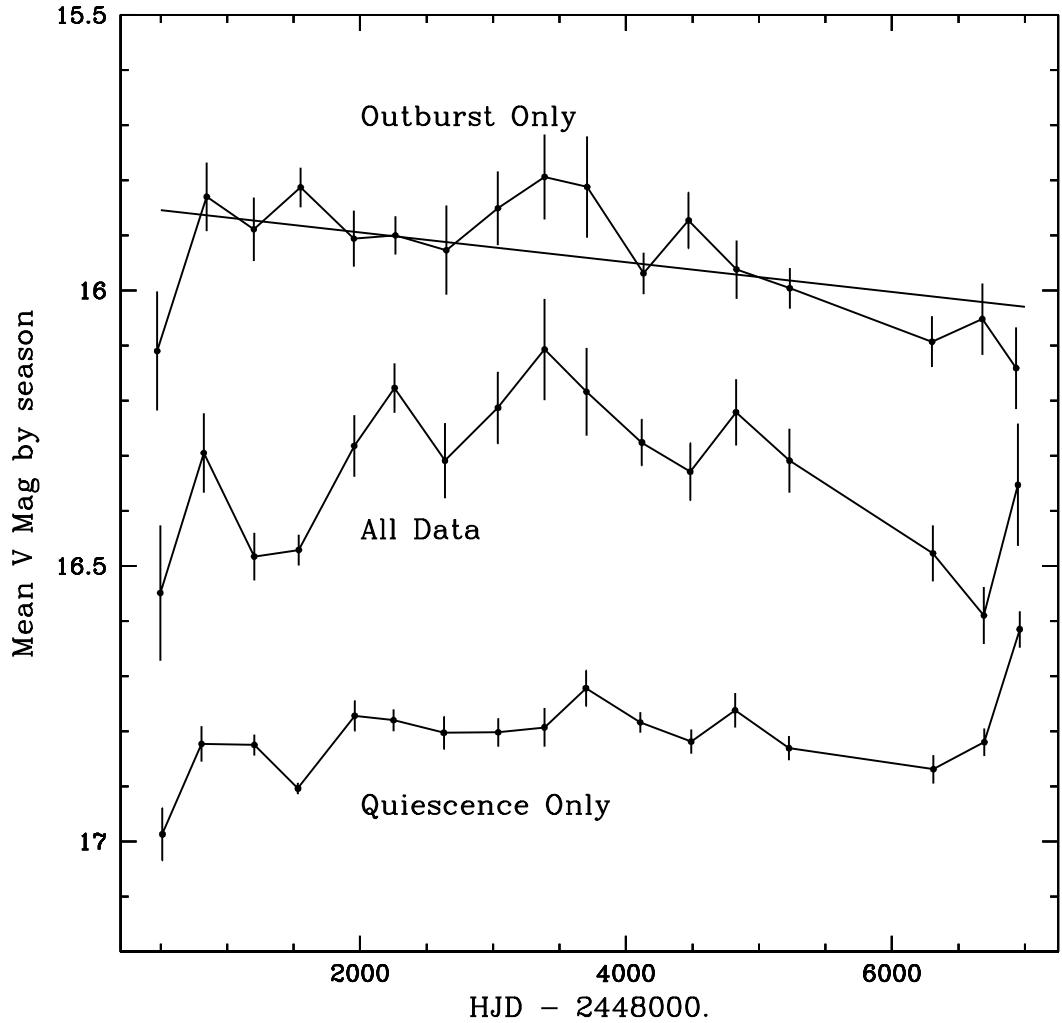}
\caption{The magnitude corresponding to the average intensity for each
observing season of V446~Her, using the data shown in Figure 1. Top is 
for outburst
data only, taken to be points with V $<$ 16.5.  Middle is for all data.
Bottom is for quiescence data only, taken to be points with V $>$ 16.5.  The
error bars are for one standard deviation of the mean.  The straight line
fit to the top plot has a slope of 0.01 mag yr$^{-1}$, consistent with
the decline of other old novae in the decades following the nova.} 
\end{figure}
\clearpage  

\begin{figure}  
\epsscale{0.9}
\plotone{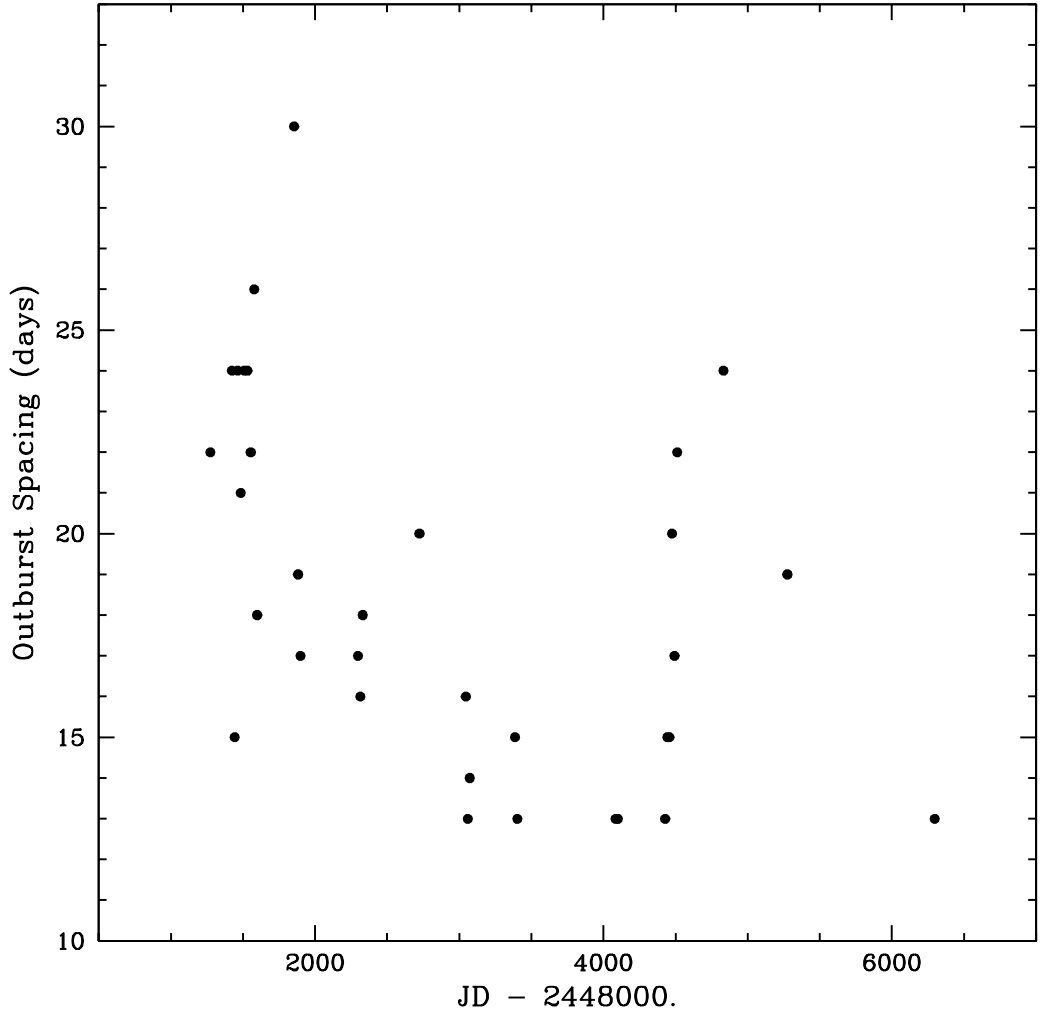}
\caption{The spacing in days of adjacent outbursts in V446~Her, over
$\sim$14 years.} 
\end{figure}
\clearpage  

\begin{figure}  
\epsscale{0.9}
\plotone{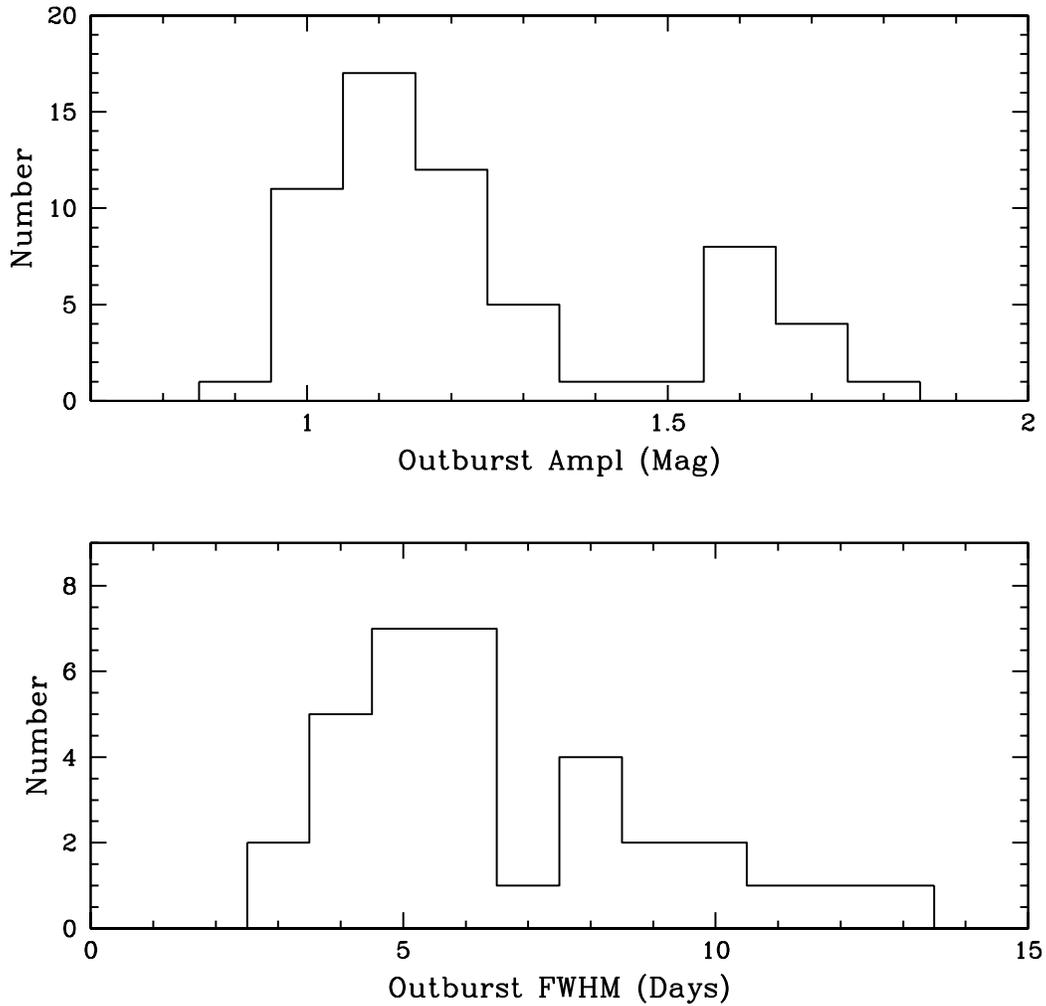}
\caption{Top: Histogram of the outburst amplitudes in V446~Her.  These 
amplitudes are attenuated by the inclusion of the constant light of the optical 
companions in the measurements (see text for details).  Bottom: Histogram
of the widths of the outbursts.  This width measurement also includes the
light of the two close companions.} 
\end{figure}
\clearpage  

\begin{figure}  
\epsscale{0.9}
\plotone{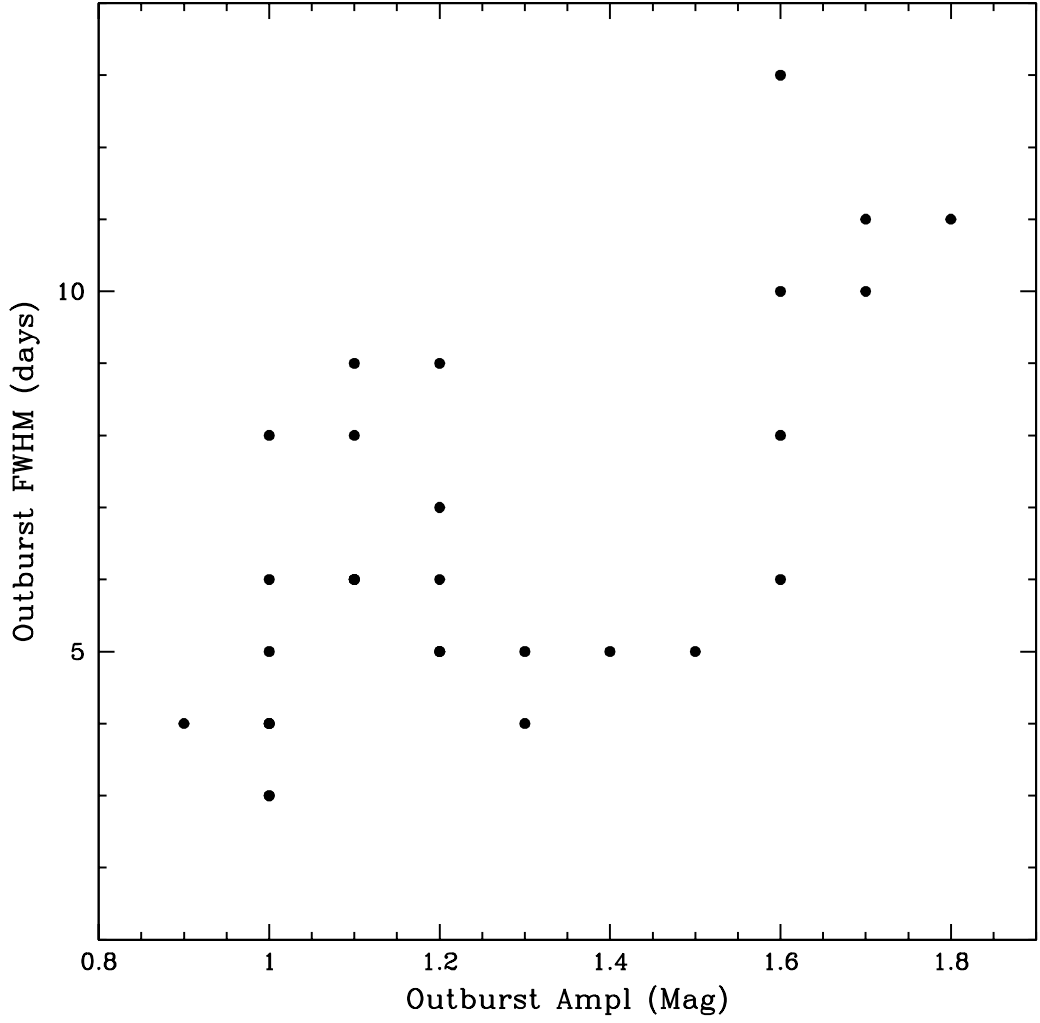}
\caption{The full-width-half-maximum of the V446~Her outbursts vs. the amplitude
of the outbursts.} 
\end{figure}
\clearpage

\end{document}